\documentclass[floatfix,showpacs,showkeys,amsmath,preprintnumbers,nofootinbib,onecolumn] {revtex4}
\usepackage{graphicx}
\usepackage{colordvi}
\usepackage{amsmath}
\usepackage{amssymb}
\usepackage{latexsym}
\usepackage{hyperref}
    \def\be{\begin{equation}}
    \def\ee{\end{equation}}
    \def\bea{\begin{eqnarray}}
    \def\eea{\end{eqnarray}}
\begin{document}
\title{Black Hole Dynamics in Power-law based Metric $f(R)$ Gravity }
\author{Suraj Kumar Pati$^a$, Bibekananda Nayak$^{a*}$ and Lambodar Prasad Singh$^b$}
\affiliation{$^a${P. G. Dept. of Applied Physics \& Ballistics, Fakir Mohan University, Balasore, Odisha-756019, India }\\
$^b${P. G. Dept. of Physics, U. N. College of Science \& Technology, Adaspur, Cuttack, Odisha-754011, India}\\
E-mail: {surrajkumarpati@gmail.com, $^*$bibekanandafm@gmail.com, lambodar\_uu@yahoo.co.in}}
\begin{abstract}
In this work, we use power-law cosmology to investigate the evolution of black holes within the context of metric $f(R)$ gravity 
satisfying the conditions provided by Starobinsky model. In our study, it is observed that presently accelerated 
expansion of the universe can be suitably explained by this integrated model without the need for dark energy. We also found that mass of a black hole decreases by
absorbing surroundings energy-matter due to modification of gravity and more the accretion rate more is mass loss. Particularly 
the black holes, whose formation masses are nearly 
$10^{20}$ gm and above, are evaporated at a particular time irrespective of their formation mass. Again our analysis reveals that 
the maximum mass of a black hole supported by metric $f(R)$ gravity is $10^{12} M_{\odot}$, where $M_{\odot}$ represents the solar mass.
\end{abstract}
\maketitle
\section{Introduction}
The recently observed accelerated expansion of the universe \cite{ae1,ae2} has put a challenge for its theoretical understanding. 
To explain it, two general ways have been used in literature. First way is by introducing a new type of energy having negative
pressure called dark energy \cite{de1,de2} and the other way is by modifying the theory of gravity \cite{fr1,
fr2}. Essentially, dark energy models modify the energy-momentum tensor associated with the matter filling the universe, 
i.e. the right hand side of Einstein$'$s equation. Whereas, modified gravity theories make a change in Einstein$'$s gravity part, 
i.e. the left hand side of the said equation. Again for addressing the long-standing horizon, flatness and monopole problems \cite{hfm1, hfm2, hfm3}, 
a phase of exponential expansion termed as inflation \cite{inf1} is thought to be occurred during early evolution of the universe. This inflation is believed to have 
provided the mechanism that generates primordial inhomogeneities, which could act as seeds for the formation of large structures \cite{lss1}. 
In between these two phases of acceleration, there must be a period of decelerated expansion during which  
primordial nucleosynthesis \cite{nucl1,nucl2} to structure formation \cite{sf1} were occurred. 

The introduction of dark energy may predict the present accelerated expansion of the
universe but its nature and characteristics are unknown. Though cosmological constant \cite{cc1,cc2,cc3,cc4} seems to be the simplest candidate for dark energy,  
but its understanding as representing the vacuum energy of quantum field, used to describe the fundamental interactions, seems to be higher than observed value by 123 orders of magnitude.
So different types of dynamical dark energy models like Quintessence \cite{qnt1,qnt2} and K-essence \cite{kes1} are proposed. 
And in some works, people also considered about an exotic form of dark energy, named as Phantom energy \cite{phan1,phan2,phan3}, which 
violates dominant energy condition. Again in literature, there are some discussion on ‘modified matter’ dynamical dark energy models like Chaplygin 
gas \cite{chap1,chap2}. But most of them are 
not able to explain all features of the universe, like for example coincidence problem: ``why the observed 
values of the cold dark matter density and dark energy density are of the same order of magnitude today although 
they differently evolve during the expansion of the universe''. Also in some works \cite{int1,int2,int3,int4,int5}, 
coupling between dark energy and dark matter 
are discussed, which can able to alleviate the coincidence problem. But till date no specific coupling in the dark sectors 
has been known, based on fundamental theories.

As an alternative to dark energy, different types of modified theories of gravity are
discussed in literature. The action for modified theories of gravity are basically extensions of
the Einstein-Hilbert action with an arbitrary function of the Ricci scalar R. These theories are
of particular interest since they naturally appear in the
low-energy effective actions of the quantum gravity \cite{qg1,qg2} and String Theory \cite{st1,st2}. In such theories both the early time 
inflation and the late-time acceleration of the universe could be
resulted by a single mechanism. Again these theories play a major role in astrophysical scales.
In fact modifying the gravity affects the gravitational potential in the low energy limit and
the modified potential reduces to the Newtonian one on the solar system scale. Moreover, a corrected gravitational 
potential could offer the possibility to fit galaxy rotation curves without the need of huge amounts of dark matter \cite{ndm1,ndm2,ndm3}.
There are many ways to modifying the theory of gravity such as $f(R)$ gravity \cite{frg1,frg2,frg3,frg4,frg5,frg6,frg7,frg8}, $f(T)$ gravity \cite{ft1,ft2}, 
$f(R,T)$ gravity \cite{frt1,frt2,frt3} etc., 
where $T$ is the trace of energy-momentum tensor.
Among them the simplest one is f(R) gravity obeying metric formalism \cite{mfr1,mfr2}. 

For studying cosmological implication of $f(R)$ models, the existence of exact power-law solutions is discussed in literature \cite{pls1,pls2} 
corresponding to phases of cosmic evolution when the energy density is dominated by a perfect fluid. The existence of such solutions 
are particularly relevant because in FRW backgrounds, they typically represent asymptotic or intermediate states in the phase space 
of the dynamical system representing all possible cosmological evolutions.

Again during the early evolution of the universe, black holes could be formed due to various mechanisms 
such as inflation \cite{cgl1,kmz1}, initial inhomogeneities \cite{carr1,swh1} and gravitational collapse \cite{khopol1,jedam1} etc. 
Literature is enriched with so many types of black holes and
among them Schwarzschild type black hole is the simplest one, 
which has no charge and no angular momentum and its formation mass could be as large as the
mass contained in the Hubble volume $m_H$ ranging down to about
$10^{-4}m_H$ \cite{bhm}. These black holes can thus span enormous mass range starting from Planck mass to
few order of solar mass. Again black holes which are formed before inflation are completely diluted 
due to exponential expansion and also environment is not suitable for them to be formed in matter-dominated era.
i.e. all the black holes are formed by the time of matter-radiation equality $t_e$. It is, generally, considered that the transition from radiation to matter-domination 
could occur when the age of the universe is nearly $10^{11}$ sec and thus the maximum formation mass of the Schwarzschild black hole would be  $(m_H)_{t_e}=G^{-1}t_e \sim 10^{49}$ gm.
But Hawking found that black holes could emit thermal radiation due to quantum effects \cite{hwk} and thus some of the black holes could be completely evaporated by present time. Those black holes, 
which are evaporated completely in the early time, 
could account for baryogenesis \cite{bar1,bar2,bar3} in the early universe. Whereas presently surviving  
black holes can act as seeds for structure formation and can also provide a significant contribution towards the
dark matter \cite{dm1,dm2,dm3,dm4,dm5}. It is also expected that black holes could absorb surroundings energy-matter during their evolution, 
which is literally known as accretion. In literature so many works \cite{acc1,acc2,acc3,acc4,acc5,acc6,acc7} are found, involving accretion of radiation, matter and dark energy. 
It is observed from those works that due to accretion of radiation, matter and vacuum energy, black holes lifetime prolongate, whereas accretion of phantom energy decreases the lifespan of black holes.

But It has been shown recently that Schwarzschild type black holes could be formed in $f(R)$ gravity, if it obeyed Starobinsky model \cite{frbh}.
The most important thing about Starobinsky model  \cite{star1,star2,star3} is that it could explain the
inflationary scenario of early Universe. In this model the Lagrangian density is taken 
as $f(R)=R+{R^2}/6{M^2}$, where $M^2$ is a phenomenological constant having dimension of $R$. 
During inflation, $R^2$ term provides a stage for
the de-Sitter-like evolution of space-time. The inflationary potential has a stable minimum,
which allows for the graceful exit, reheating and good low-energy limit of the theory. 

In this work, we use power-law cosmology for studying the evolution of Schwarzschild type black holes in Starobinsky type metric $f(R)$ gravity. 
We, here, assume that after inflation, the universe witnessed radiation-dominated era and then finally matter-dominated era.
In our analysis, first, we show that how the density of the energy-matter filling the universe get changed due to modification of gravity and 
then try to explain the accelerated expansion of the universe in terms of it. Finally, we discuss the evolution of the black holes in this environment.

\section{Basic Framework}
For metric f(R) gravity, the action can be written as \cite{fr1}
\be \label{1}
S=\frac{1}{16 \pi G}\int \sqrt{-g} f(R)d^4x+S_M,
\ee
where $S_M$ is the action due to non-gravitational part of the universe.
From the variation of the metric, equation (\ref{1}) yields the field equation
\be \label{2}
f^\prime (R) R_{\mu\nu}-\frac {1}{2}f(R)g_{\mu\nu}+(g_{\mu\nu}\square-{\nabla_\mu} {\nabla_\nu})f^\prime (R) =8 \pi G T_{\mu\nu},
\ee
where the prime denotes differentiation with respect to R, $R_{\mu\nu}$ is the Ricci tensor, R is the
Ricci scalar, $g_{\mu\nu}$ is the metric tensor, $\nabla _\alpha$ and $\nabla_\beta$ are the covariant derivative of the metric
tensor, $\square=g^{\alpha\beta}\nabla_\alpha \nabla_\beta$ is the d’Alembert operator and $T_{\mu\nu}$ is the energy-momentum tensor. 
Equation (\ref{2}) is a fourth-order partial differential equation in the metric since $R$ already includes its second derivative. 
For an action that is linear in $R$, the fourth order terms (the last term of the left hand side of equation (\ref{2})) vanish and 
the theory reduces to General Theory of Relativity.

Here we consider that the homogeneous and isotropic universe is described by the
Friedmann-Robertson-Walker (FRW) metric
\be \label{3}
d\tau^2=-dt^2+a^2(t)\left[{\frac{dr^2}{1-kr^2}}+{r^2}d{\theta^2}+{r^2}{sin^2}\theta{d\phi^2}\right],
\ee
with $a(t)$ as the scale factor and $k$ as the spatial curvature having values $+1$, $0$ and $-1$ for closed, flat and open universe respectively. 

So from field equation (\ref{2}), one can get Friedmann equations for a spatially flat FRW universe ($k=0$) as
\bea \label{4}
H^2=\frac{1}{3f^\prime} \Big[8\pi G \rho + \frac{1}{2} (R f^\prime-f)-3 H\dot{R}f^{\prime \prime} \Big],
\eea
and 
\bea \label{5}
2\dot{H}+3H^2=-\frac{1}{f^\prime} \Big[8 \pi G p+ {{\dot{R}}^2}f^{\prime \prime \prime}+2 H \dot{R} f^{\prime \prime}+ \ddot{R} f^{\prime \prime} + \frac{1}{2}(f-R f^{\prime}) \Big],
\eea
where $f^{\prime}=\frac{\partial f(R)}{\partial R}$, $f^{\prime \prime}=\frac{\partial^2 f(R)}{\partial R^2}$, 
$f^{\prime \prime \prime}=\frac{\partial^3 f(R)}{\partial R^3}$, $H=\frac{\dot{a}}{a}$ is the Hubble parameter 
and $\dot{H}=\frac{dH}{dt}$, $\rho$ and  $p$ are the density and pressure of the perfect fluid filling the universe 
and connected by the equation of state $p= \gamma \rho$. Here it is assumed that $f^{\prime}> 0$ in order to have a 
positive effective gravitational coupling and $f^{\prime \prime}>0$ to fulfil the requirements of stability of the classical solutions of the Einstein equation.\\
Also from FRW metric as given in (\ref{3}), we found the value of Ricci scalar$(R)$ as
\be \label{6}
R=12 H^2 + 6 \dot{H}.
\ee

The energy conservation equation then becomes
\be \label{7}
\dot{\rho}+3 H (1+\gamma) \rho=0,
\ee
which implies $\rho \propto a^{-3(1+\gamma)}$, where $\gamma$ is equation of state parameter having values $\frac{1}{3}$ for radiation-dominated era and $0$ for matter-dominated era. 
Again Using equations (\ref{4}) and (\ref{5}), one can get the Raychaudhuri equation as
\bea \label{8}
\dot{H}=-\frac{1}{2 f^\prime} \Big[8 \pi G (1+\gamma) \rho+ {{\dot{R}}^2}f^{\prime \prime \prime}-H \dot{R} f^{\prime \prime}+ \ddot{R} f^{\prime \prime} \Big].
\eea
\section{Starobinsky Model and Power-law Cosmology}
In our work, we choose Starobinsky model \cite{star1,star2,star3}, where
\be \label{9}
f(R)=R+\frac{R^2}{6M^2}.
\ee

Now by differentiating $f(R)$ with respect to R, we can get
$f^\prime =1+\frac{R}{3M^2}$, $f^{\prime \prime}=\frac{1}{3M^2}$ and $f^{\prime \prime \prime}=0$.
\\Here $M^2$ is a phenomenological constant having dimension of $R$. \\
\\By using the above values of $f(R)$ and its derivatives in equation (\ref{4}), we found
\bea \label{10}
\rho=\frac{1}{8 \pi G}\Big[3\Big(1+\frac{R}{3M^2}\Big)H^2-\frac{R^2}{12 M^2}+\frac{H \dot{R}}{M^2} \Big].
\eea
Again taking the values of $R$ from equation (\ref{6}), equation (\ref{10}) can be written as
\bea \label{11}
\rho=\frac{1}{8 \pi G M^2}\Big[3M^2 H^2+18 \dot{H} H^2+6 H \ddot{H}-3 \dot{H}^2 \Big].
\eea
But in general power law cosmology, the scale factor varies with time as a power law, i.e. $a(t) \propto t^{\beta}$. So the Hubble parameter and its derivatives become
$H=\frac{\beta}{t}$, $\dot{H}=-\frac{\beta}{t^2}$, $\ddot{H}=\frac{2 \beta}{t^3}$ and $\dddot{H}=-\frac{6}{t^4}$.\\
Now on simplification, equation (\ref{11}) becomes
\bea \label{12}
\rho=\frac{3 \beta^2}{8 \pi G t^2}\Big[1-\frac{3(2 \beta-1)}{M^2 t^2} \Big],
\eea
which gives
\bea \label{13}
\frac{\dot{\rho}}{\rho}=-\frac{1}{t}\Big[\frac{2M^2t^2+12-24\beta}{M^2t^2+3-6\beta}\Big].
\eea
But using general power-law concept in energy conservation equation (\ref{7}), we get
\be \label{14}
\rho \propto t^{-3\beta(1+\gamma)},
\ee
which implies
\bea \label{15}
\frac{\dot{\rho}}{\rho}=-\frac{3\beta(1+\gamma)}{t}.
\eea
Comparing equations (\ref{13}) and (\ref{15}), one can find
\bea \label{16}
\beta(1+\gamma)=\frac{2}{3}\Big[\frac{M^2t^2+6-12\beta}{M^2t^2+3-6\beta}\Big].
\eea
Like standard model of cosmology, here we consider that present universe is matter-dominated ($\gamma=0$) and before it was radiation-dominated ($\gamma=\frac{1}{3}$).\\
Now for radiation-dominated era, equation (\ref{16}) gives
\bea \label{17}
12 \beta^2 - (18+2 M^2t^2) \beta +(6+M^2t^2)=0.
\eea
The solutions of above equation (\ref{17}) are $\beta=\frac{1}{2}$ and $\beta=1+\frac{M^2t^2}{6}$. But $\beta=1+\frac{M^2t^2}{6}$ is prohibited, since it makes the density of energy-matter 
filling the universe negative. So like standard model of cosmology and scalar-tensor theory \cite{bc}, here also scale factor varies in radiation-dominated era as
\be \label{18}
a(t) \propto t^{\frac{1}{2}},
\ee
which has a strong observational support.\\
Again for matter-dominated era, equation (\ref{16}) gives
\bea \label{19}
18 \beta^2 - (33+3 M^2t^2) \beta +(12+2 M^2t^2)=0.
\eea
The solution of above equation (\ref{19}) are $\beta=\frac{1}{12}\Big[(11+M^2t^2)+\sqrt{(5+M^2t^2)^2-4M^2t^2}\Big]$ and 
$\beta=\frac{1}{12}\Big[(11+M^2t^2)-\sqrt{(5+M^2t^2)^2-4M^2t^2}\Big]$.
But the root $\beta=\frac{1}{12}\Big[(11+M^2t^2)-\sqrt{(5+M^2t^2)^2-4M^2t^2}\Big]$, which approaches standard model 
value $\beta=\frac{2}{3}$ in the limit $\frac{1}{M^2t^2} \rightarrow 0$, is not suitable for providing presently observed accelerated expansion. 
Thus in matter-dominated era scale factor varies like
\bea \label{20}
a(t) \propto t^{\frac{1}{12}\Big[(11+M^2t^2)+\sqrt{(5+M^2t^2)^2-4M^2t^2}\Big]}.
\eea

\section{Deceleration Parameter}
Though Hubble’s law is successful in explaining the expansion of the universe, but it is unable to predict the nature of the expansion. 
The cosmological term which can be used to determine the nature of expansion is known as deceleration parameter and its mathematical form is
\be \label{21}
q=-\frac {\ddot a(t) a(t)}{[\dot a(t)]^2},
\ee
which on use of equation (\ref{8}) gives
\bea \label{22}
q=-1+\frac{1}{2 f^\prime H^2} \Big[8 \pi G (1+\gamma) \rho+ {{\dot{R}}^2}f^{\prime \prime \prime}-H \dot{R} f^{\prime \prime}+ \ddot{R} f^{\prime \prime} \Big].
\eea
Now above equation (\ref{22}) can be simplified as
\bea \label{23}
q=-1+\frac{4 \pi G \rho (1+\gamma) M^2 t^4 + 4 \beta^3+ 10 \beta^2 -6 \beta}{4 \beta^4-2 \beta^3 +M^2 t^2 \beta ^2},
\eea
which implies for $t \rightarrow 0$ and  large value of $\beta$, $q \approx -1 + \frac{1}{\beta} \approx -1$. This supports the idea of inflation that the universe undergoes a phase of exponential expansion during early period of evolution. Because for exponential expansion $a(t) \propto e^{\alpha t}$ which, in turn, gives $q=-\frac{\ddot{a}(t) a(t)}{{\dot{a}(t)}^2}=-1$. 
 
For radiation-dominated era, the equation (\ref{23}) gives
\bea \label{24}
q=-1+\frac{64 \pi G \rho t^2}{3}.
\eea
Substituting the value of $\rho$ from equation (\ref{12}) with $\beta=\frac{1}{2}$ in equation (\ref{24}), the deceleration parameter for radiation-dominated era is found to be $q=1$. This provides decelerated expansion through out the radiation-dominated era.

But for matter-dominated era, using equations (\ref{23}) and (\ref{12}), we get
\bea \label{25}
q=-1+\frac{\frac{3}{2} M^2t^2 \beta^2 -5\beta^3+\frac{29}{2}\beta^2-6\beta}{4\beta^4-2\beta^3+M^2t^2 \beta^2},
\eea
where $\beta=\frac{1}{12}\Big[(11+M^2t^2)+\sqrt{(5+M^2t^2)^2-4M^2t^2}\Big]$.\\
Solving above equation (\ref{25}) by taking present age of the universe ($t_0$) as $13.82 \times 10^9$ years, we construct the Table-I for presenting the variation of present value of deceleration parameter $q_0$ with $M^2{t_0}^2$.

\begin{table}[ht] 
\center
\caption{The variation of present value of deceleration parameter $(q_0)$ with $M^2{t_0}^2$ is given in the Table.}
\begin{tabular}{|c| c ||c| c |}\hline
 \hspace{0.2 in} $q_0$ \hspace{0.2 in} & \hspace{0.2 in} $M^2{t_0}^2$ \hspace{0.2 in} & \hspace{0.2 in} $q_0$ \hspace{0.2 in} & \hspace{0.2 in} $M^2{t_0}^2$ \hspace{0.2 in}\\\hline 
-0.30 & 0.696 & -0.65 & 9.838\\
-0.35 & 1.466 & -0.70 & 12.75\\
-0.40 & 2.333 & -0.75 & 16.80\\
-0.45 & 3.331 & -0.80 & 22.846\\
-0.50 & 4.501 & -0.85 & 32.889\\
-0.55 & 5.905 & -0.90 & 52.929\\
-0.60 & 7.636 & -0.90 & 112.966\\\hline
\end{tabular}
\end{table}

Comparing Table-I with the observational data \cite{dec1} that $q_0 \approx -0.55$, we take $M^2{t_0}^2=5.905$ for the rest part of the paper.\\
Now the evolution of scale factor in matter-dominated era due to modification of gravity can be picturized from equation (\ref{19}), which is shown in Figure-1.

\begin{figure}[h] 
\centering
\includegraphics[scale=0.8]{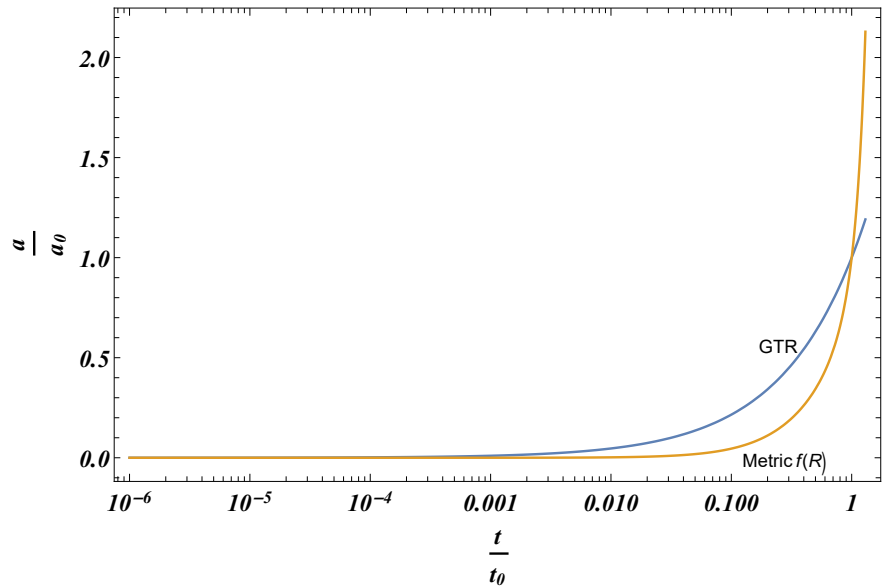} 
\caption{Evolution of scale factors in matter-dominated era due to normal gravity and metric $f(R)$ gravity}
\end{figure}

\section{Evolution of Black Holes}
During their evolution, black holes are affected both by Hawking evaporation and accretion of surroundings energy matter. \\
The mass of a black hole can be decreased due to Hawking evaporation by obeying the formula \cite{hwk,bc}
\bea \label{26}
\dot{m}= -\frac{a_H}{256 \pi^3} \frac{1}{G^2m^2},
\eea
where $a_H$ is the black body constant and $G$ is Newton's gravitational constant.

Again black holes mass could be changed due to the accretion of energy-matter from the surroundings, when the gravity is modified due to metric formalism. 
As expected, this accretion rate is proportional to the product of the surface area of the black hole and the surrounding density and this accretion rate 
can be controlled by introducing a controlling factor like accretion efficiency $f$. 
Thus in metric $f(R)$ gravity the absorption of surroundings energy-matter could change the mass of a black hole as \cite{acc3}
\bea \label{27}
\dot{m}=4 \pi r^2_{bh} f \rho. 
\eea
But the radius of Schwarzschild black hole is  $r_{bh}=2Gm$.
So by using equation (\ref{12}) in equation (\ref{27}), we get
\bea \label{28}
\dot{m}=\frac{6 G f m^2 \beta^2}{t^2} \Big[1+\frac{3-6 \beta}{M^2t^2} \Big].
\eea

In metric $f(R)$ gravity, thus, the complete evolution equation of a black hole becomes
\bea \label{29}
\dot{m}= -\frac{a_H}{256 \pi^3} \frac{1}{G^2m^2}+\frac{6 G f m^2 \beta^2}{t^2} \Big[1+\frac{3-6 \beta}{M^2t^2} \Big].
\eea
Now the evolution of black holes can be studied considering two epochs separately.

\subsection{Radiation-dominated era}
Using the value of $\beta$ as $\frac{1}{2}$ in accretion equation (\ref{27}), we get
\bea \label{30}
\dot{m}=\frac{3}{2}G f_{rad} \frac{m^2}{t^2}.
\eea
The solution of above differential equation (\ref{30}) can be written as
\bea \label{31}
m=m_i\Big[1+\frac{3}{2}f_{rad}(\frac{t_i}{t}-1) \Big]^{-1},
\eea 
where $t_i$ and $m_i$ are the formation time and formation mass of a black hole respectively. 
For large time $t$, the above equation asymptotes to $m=\frac{m_i}{1-\frac{3}{2}f_{rad}}$. 
Thus for accretion to be effective $f_{rad} < \frac{2}{3}$.\\
Now the complete evolution equation of a black hole in radiation-dominated era becomes 
\bea \label{32}
\dot{m}= -\frac{a_H}{256 \pi^3} \frac{1}{G^2m^2}+\frac{3}{2}G f_{rad} \frac{m^2}{t^2},
\eea
which is same as the standard model of cosmology \cite{pra}.

\subsection{Matter dominated era}
Since black holes, in general, can not be formed in matter-dominated era, 
here we discuss the evolution of those black holes which are already formed during radiation-dominated era. \\
The accretion equation (\ref{27}), in matter-dominated era, has the form
\bea \label{33}
\dot{m}=\frac{G f_{mat} m^2}{24 t^2} \Big(\frac{1}{M^2t^2}\Big) \Big( \frac{M^2t^2}{2}-\sqrt{(5+M^2t^2)^2-4M^2t^2}-\frac{5}{2} \Big)  \nonumber \\
\Big[(11+M^2t^2)+\sqrt{(5+M^2t^2)^2-4M^2t^2}\Big]^2.
\eea
Using numerical solution of equations (\ref{33}), we plot Figure-2 which shows
the variation of black holes' mass with time in metric $f(R)$ gravity.

\begin{figure}[h] 
\centering
\includegraphics[scale=0.8]{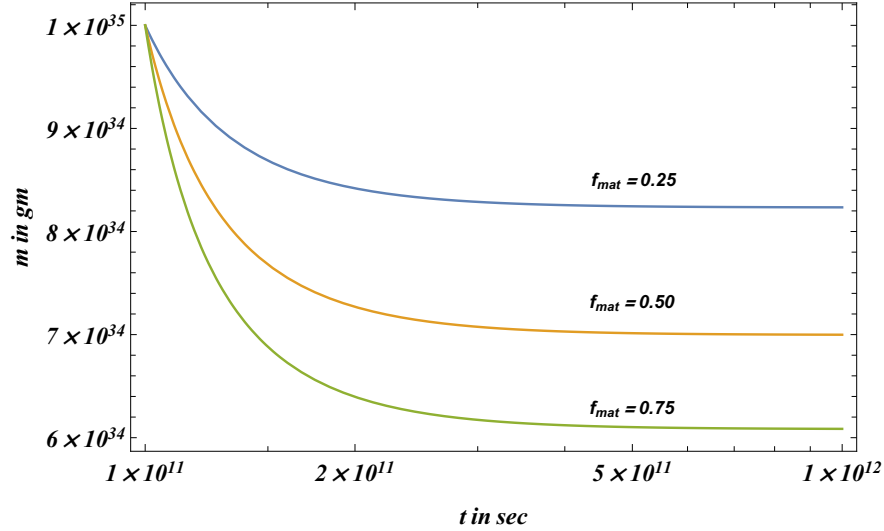}
\caption{Variation of black holes' mass with time having formation mass $10^{38}$ gm by considering only accretion in matter-dominated era.}
\end{figure} 

From Figure-2, it is evident that the mass of a black hole decreases due to accretion of surroundings energy-matter in matter-dominated era. 
This, we believe, is an interesting result being counter-intuitive. 

Now the complete evolution equation of a black hole, in matter-dominated era, becomes 
\bea \label{34}
\dot{m}=-\frac{a_H}{256 \pi^3} \frac{1}{G^2m^2}+\frac{G f_{mat} m^2}{24 t^2} 
\Big(\frac{1}{M^2t^2}\Big) 
\Big( \frac{M^2t^2}{2}-\sqrt{(5+M^2t^2)^2-4M^2t^2}-\frac{5}{2} \Big) \nonumber \\
\Big[(11+M^2t^2)+\sqrt{(5+M^2t^2)^2-4M^2t^2}\Big]^2 , 
\eea
which implies that the accretion term would be effective, if the black holes mass 
at the time of matter-radiation equality ($m(t_e)$) satisfies the condition
\bea \label{35}
m(t_e) > \Big|\frac{a_H}{256 \pi^3}\frac{24 {t_e}^2}{G^3 f_{mat}} (M^2{t_e}^2) \Big( \frac{M^2{t_e}^2}{2}-\sqrt{(5+M^2{t_e}^2)^2-4M^2{t_e}^2}-\frac{5}{2} \Big)^{-1} \nonumber\\ 
\Big[(11+M^2{t_e}^2)+\sqrt{(5+M^2{t_e}^2)^2-4M^2{t_e}^2}\Big]^{-2}\big|^{\frac{1}{4}} . 
\eea
Since $M^2{t_0}^2=5.905$ and $0<f_{mat}<1$, on simple calculation above equation (\ref{35}) predicts that those black holes would be affected by accretion, 
whose masses at the time of matter-radiation equality $M(t_e)$ are greater that $10^{19}$ gm.\\
Solving the equations (\ref{32}) and (\ref{34}) numerically, we
 construct Table-II, where the variation of evaporation times of black holes with rate of accretion of surroundings energy-matter in the presence of metric $f(R)$ gravity  
is presented. (The subscript $i$ refers to the initial value.)

\begin{table}[ht] 
\center
\caption{The  evaporation times of black holes in metric f(R) gravity for $f_{rad}=0.6$ }
\begin{tabular}{|c|c|c|c|c|c|c|c|}\hline
$t_i$  & $m_i$   & $({t_{evap}})_{GTR}$  & $(t_{evap})_{fr}$ & $(t_{evap})_{fr}$ & $(t_{evap})_{fr}$ & $(t_{evap})_{fr}$ & $(t_{evap})_{fr}$\\
(in sec) & (in gm) & (in sec) & (in sec) & (in sec) & (in sec) & (in sec) & (in sec) \\
 & & $f_{mat}=0$ &  $f_{mat}=0.2$ & $f_{mat}=0.4$ & $f_{mat}=0.6$ & $f_{mat}=0.8$ & $f_{mat}=1.0$ \\\hline
$10^{-23}$ & $10^{15}$ & $3.33 \times10^{19}$ & $3.33 \times10^{19}$ & $3.33 \times10^{19}$ & $3.33 \times10^{19}$ & $3.33 \times10^{19}$ & $3.33 \times10^{19}$ \\
$10^{-21}$ & $10^{17}$ & $3.33 \times10^{25}$ & $3.33 \times10^{25}$ & $3.33 \times10^{25}$ & $3.33 \times10^{25}$ & $3.33 \times10^{25}$ & $3.33 \times10^{25}$ \\
$10^{-19}$ & $10^{19}$ & $3.33 \times10^{31}$ & $1.02 \times10^{30}$ & $8.37 \times10^{29}$ & $7.43 \times10^{29}$ & $6.83 \times10^{29}$ & $6.39 \times10^{29}$ \\
$10^{-18}$ & $10^{20}$ & $3.33 \times10^{34}$ & $1.09 \times10^{30}$ & $8.86 \times10^{29}$ & $7.85 \times10^{29}$ & $7.20 \times10^{29}$ & $6.74 \times10^{29}$ \\
$10^{-17}$ & $10^{21}$ & $3.33 \times10^{37}$ & $1.10 \times10^{30}$ & $8.91 \times10^{29}$ & $7.89 \times10^{29}$ & $7.24 \times10^{29}$ & $6.77 \times10^{29}$ \\
$10^{-13}$ & $10^{25}$ & $3.33 \times10^{49}$ & $1.10 \times10^{30}$ & $8.91 \times10^{29}$ & $7.89 \times10^{29}$ & $7.24 \times10^{29}$ & $6.77 \times10^{29}$ \\
$10^{-8}$  & $10^{30}$ & $3.33 \times10^{64}$ & $1.10 \times10^{30}$ & $8.91 \times10^{29}$ & $7.89 \times10^{29}$ & $7.24 \times10^{29}$ & $6.77 \times10^{29}$ \\
$10^{-3}$  & $10^{35}$ & $3.33 \times10^{79}$ & $1.10 \times10^{30}$ & $8.91 \times10^{29}$ & $7.89 \times10^{29}$ & $7.24 \times10^{29}$ & $6.77 \times10^{29}$ \\
$10^{0}$   & $10^{38}$ & $3.33 \times10^{88}$ & $1.10 \times10^{30}$ & $8.91 \times10^{29}$ & $7.89 \times10^{29}$ & $7.24 \times10^{29}$ & $6.77 \times10^{29}$ \\
$10^{2}$   & $10^{40}$ & $3.33 \times10^{94}$ & $1.10 \times10^{30}$ & $8.91 \times10^{29}$ & $7.89 \times10^{29}$ & $7.24 \times10^{29}$ & $6.77 \times10^{29}$ \\
$10^{6}$   & $10^{44}$ & $3.33 \times10^{106}$ & $1.10 \times10^{30}$ & $8.91 \times10^{29}$ & $7.89 \times10^{29}$ & $7.24 \times10^{29}$ & $6.77 \times10^{29}$ \\
$10^{7}$   & $10^{45}$ & $3.33 \times10^{109}$ & $1.10 \times10^{30}$ & $8.91 \times10^{29}$ & ----- & ----- & ----- \\\hline
\end{tabular}
\end{table}

From Table-II, it is found that the modification of gravity enhances the evaporation of black holes. Again all those black holes, having formation masses greater
than nearly $10^{20}$ gm, would be evaporated at a particular time depending on their accretion efficiency only. But the lifetimes of black holes, which are evaporated by now, are 
not affected by the metric $f(R)$ gravity.

\section{Discussion and Conclusion}
In this paper, we have used the metric $f(R)$ gravity obeying Starobinsky model for investigating the black hole dynamics. We, here, assume that after inflation, 
the evolution of the universe was occurred through conventional cosmological eras of radiation-domination and matter-domination basing on power-law cosmology. 
We first evaluated the modified density of the energy-matter filling the universe and then determined the scale factor for both 
the eras. From these calculations, we found that in radiation-dominated era, the evolution of the universe remains same as in the case of 
standard model of cosmology and scalar-tensor theory. 
Hence our integrated modified gravity model does not affect the early observational facts, starting from primordial nucleosynthesis 
to providing the stage for formation of large scale structures. But in matter-dominated era, modified gravity plays its role by affecting the evolution of the universe.  
Assuming that present universe is matter-dominated, we are successful in explaining the observed accelerated expansion of the universe without requiring the dark energy. 
We, finally, discuss the black hole dynamics within the context of our integrated modified gravity model. From our analysis, it is observed that black hole mass decreases due to accretion of 
surroundings energy-matter in metric $f(R)$ gravity and more the accretion rate more is mass loss. Particularly, black holes, whose masses at the time of 
matter-radiation equality $M(t_e)$ are greater that $10^{19}$ gm, would be affected by this modified gravity theory. So in addition with Hawking evaporation,  
accretion also helps in decreasing the masses of those black holes and hence those black holes will be evaporated at a quicker rate in metric $f(R)$ gravity, 
in comparison with that of standard cosmology and scalar-tensor theory.  
Moreover, the black holes, having formation masses $10^{20}$ gm and more, will show a peculiar behaviour. They all will be evaporated at a particular time 
irrespective of their formation mass depending only on rate of accretion. This can be explained by 
the fact that by that time the modification of gravity will touch its saturation point so that the universe will show  phantom type behavior \cite{epjc,bnn}. 
Also in case of modified gravity, we found that the evolution of presently evaporating black holes, whose formation masses ($M_i$) are of the order of $ 10^{15}$ gm, 
could not be disturbed by the accretion of energy-matter during matter-dominated era. 
Thus, assuming metric $f(R)$ gravity as the theory of gravity, in place of general theory of relativity, would not alter the observed astrophysical constraints \cite{jcap} on black holes. 
Again our analysis reveals that the maximum mass of a black hole supported 
by metric $f(R)$ gravity is $10^{12} M_{\odot}$, where $M_{\odot} \approx 2 \times 10^{33}$ gm is the solar mass.
\\
\\
\\
\noindent{\bf{Acknowledgement}}\\
One of our authors Bibekananda Nayak acknowledges the support from the UGC Start-Up-Grant Project having Letter No. F. 30-390/2017 (BSR) of University Grants Commission, New Delhi.

\end{document}